\documentclass[showpacs,showkeys,eqsecnum,twocolumn]{revtex4}

\usepackage{amsfonts,amsmath,amssymb,bm,graphicx,hyperref}

\begin{document}

\preprint{HUPD 0603}
%\preprint{E-print archive: hep-ph/0601156}

\title{Evolution of coupled fermions under
the influence of \\ an external axial-vector field}

\author{Maxim Dvornikov}
\affiliation{Institute of Terrestrial Magnetism,
Ionosphere \\ and Radiowave Propagation (IZMIRAN) \\
142190, Troitsk, Moscow region, Russia} \email{maxdvo@izmiran.ru}
\affiliation{Graduate School of Science, Hiroshima University,
Higashi-Hiroshima, Japan}

\date{\today}

\begin{abstract}
The evolution of coupled fermions interacting with external
axial-vector fields is described with help of the classical field
theory. We formulate the initial conditions problem for the system
of two coupled fermions in (3+1)-dimensional space-time. This
problem is solved using the perturbation theory. We obtain in the
explicit form the expressions for the leading and next to the
leading order terms in the expansion over the strength of external
fields. It is shown that in the relativistic limit the intensity
of the fermion field coincides with the transition probability in
the two neutrinos system interacting with moving and polarized
matter.
\end{abstract}



\maketitle

\section{Introduction}

%The description of the mixed fermions evolution attracts
%considerable attention especially after the experimental
%confirmation of solar neutrinos oscillations (see, e.g.,
%Refs~\cite{Aha04,Hos05}). It is well known that a mixing between
%different neutrino flavors is the essential point for the
%possibility of the neutrino conversion from one type into another.
%According to up-to-date experimental results the disappearance of
%the solar electron neutrinos can be explained by the
%Mikheyev-Smirnov-Wolfenstein (MSW) solution in supposition of the
%large vacuum mixing angle. In this paper we mainly focus on the
%mixing between different neutrino flavor eigenstates.

The description of the mixed fermions evolution attracts
considerable attention after the experimental confirmation of
solar neutrino oscillations \cite{Aha04,Hos05}. The majority of
the neutrino oscillations studies involve the quantum mechanical
approach to the description of the neutrino wave function
evolution (see especially Ref.~\cite{Kay81}). Despite of the fact
that quantum mechanics allows one to establish the main properties
of the neutrino oscillations process this method of the neutrino
oscillations treatment has several disadvantages. Neutrinos are
usually supposed to be scalar particles without reference to the
multicomponent single neutrino wave function. The famous
Pontecorvo formula (see Ref.~\cite{Pon58eng}), which is in use in
many theoretical and experimental studies of neutrino
oscillations, is valid only for ultrarelativistic particles.
However, the exact theory of the considered process must be
applicable for neutrinos with arbitrary energies. Quantum
mechanical approach also does not make it clear if mass or flavor
eigenstates bear more physical meaning. Therefore one can see that
a theoretical model of neutrino oscillations, which would overcome
the mentioned above difficulties, should be put forward. There
were numerous attempts to construct the appropriate formalism for
neutrino flavor oscillations in vacuum. The quantum field theory
was applied to this problem in
Refs.~\cite{GiuKimLeeLee93,BlaVit95,GriSto96,Beu02}. The authors
of these papers reproduced the Pontecorvo formula and discussed
the corrections to this expression. Recently we revealed in
Ref.~\cite{Dvo05} that neutrino flavor oscillations in vacuum
could be explained in frames of the classical field theory.

It was also realized that neutrino interactions with external
fields can drastically change the picture of the oscillations
process. For example, it was discovered in
Refs.~\cite{Wol78,MikSmi85eng} that the transition probability can
achieve great values if a neutrino interacts with background
matter by means of weak currents. Thus we should develop now not
only the appropriate theory of neutrino oscillations in vacuum,
but also include in out treatment possible effects of neutrino
interactions with external fields. During the last three decades
the approaches for the theoretical substantiation of the
Mikheyev-Smirnov-Wolfenstein (MSW) effect were developed. Among
them we can distinguish Refs.~\cite{Man88,Pan92} in which the
neutrino relativistic wave equations in dense matter were
analyzed. The $S$-matrix approach was used Ref.~\cite{CarChu99} to
account for the MSW effect. The influence of moving and polarized
matter was described in Refs.~\cite{NunSemSmiVal97,GriLobStu02}.

The purpose of the present work is to provide deeper understanding
of the neutrino flavor oscillations phenomenon. The approach
developed in our paper can not only reproduce the Pontecorvo
formula for the transition probability, but also give clear
physical explanation of the corrections to this expression, which
are widely discussed now (see, e.g., review~\cite{Beu03} and
references therein). The analysis used in this article is based on
classical field theory methods. We study the evolution of coupled
classical fermions under the influence of external axial-vector
fields. \emph{A classical fermion} is regarded as a first
quantized field because the Dirac equation,
\begin{equation*}%\label{Dirac}
  \frac{1}{c}
  \frac{\partial\psi}{\partial t}+
  \bm{\alpha}\bm{\nabla}\psi+
  \frac{imc}{\hbar}\beta\psi=0,
\end{equation*}
in which we use the common notations for the gamma matrices
$\bm{\alpha}=\gamma^0\bm{\gamma}$ and $\beta=\gamma^0$, does
contain Plank constant $\hbar$, however the wave function $\psi$
is supposed to be the non-operator object in our approach.
Therefore we do not involve the second quantization in the present
work. Note that the discussion of the first quantized neutrino
fields is also presented in Ref.~\cite{Nis05}.

In Sec.~\ref{GF} we start from the flavor neutrino Lagrangian
which accounts for the interaction with external axial-vector
fields. Then we derive the basic integro-differential equations
for the "mass eigenstates" which exactly take into account both
Lorentz invariance and the interaction with external fields. These
equations are also valid in (3+1) dimensions. The perturbation
theory is used for the analysis of the obtained equations. In
Sec.~\ref{VACUUM} we discuss the neutrino fields distributions and
obtain the zero order term in their expansion over the external
fields strength. This result corresponds to vacuum neutrino
oscillations. In Sec.~\ref{FIELD} we get the first order
correction to the neutrino field intensity in vacuum and show that
our formula is identical to the expression for the neutrino
transition probability obtained in the quantum mechanical approach
for ultrarelativistic neutrinos. This case corresponds to neutrino
flavor oscillations in moving and polarized matter. Finally we
discuss our results in Sec.~\ref{CONCL}.

\section{General formalism}\label{GF}

Without losing generality we can discuss the evolution of the two
coupled fermions system $(\nu_1,\nu_2)$. These fermions are taken
to interact with the external axial-vector fields $f_{1,2}^\mu$.
The Lagrangian for this system is expressed in the following form
\begin{align}\label{Lagrnu}
  \mathcal{L}(\nu_1,\nu_2)= & \sum_{k=1,2}\mathcal{L}_0(\nu_k)+
  g\bar{\nu}_2\nu_1+g^*\bar{\nu}_1\nu_2
  \notag \\
  & -
  \sum_{k=1,2}
  \bar{\nu}_k\gamma_\mu^L\nu_k f_k^{\mu},
\end{align}
where $g$ is the coupling constant ($g^*$ is the complex conjugate
value), $\gamma_\mu^L=\gamma_\mu(1+\gamma^5)/2$
($\gamma^5=-i\gamma^0\gamma^1\gamma^2\gamma^3$), and
\begin{equation*}
  \mathcal{L}_0(\nu_k)=
  \bar{\nu}_k(i\gamma^\mu\partial_\mu-\mathfrak{m}_k)\nu_k.
\end{equation*}
is the free fermion Lagrangian, $\mathfrak{m}_k$ are the masses of
the corresponding fermions $\nu_k$.

One of the possible examples of the fermions $\nu_k$ is the system
of neutrinos belonging to different flavor states. In this case we
can identify the first fermion in Eq.~\eqref{Lagrnu} with a muon
neutrino $\nu_{\mu}$ or a $\tau$-neutrino $\nu_{\tau}$ and the
second one with an electron neutrino $\nu_{e}$. These neutrino
types are known to interact with matter composed of electrons,
protons and neutrons by means of the electroweak interactions.
Note that an electron neutrino interacts with background fermions
via both charged and neutral weak currents whereas a muon or a
$\tau$-neutrino are involved only in the interaction through weak
neutral currents. Thus the external axial-vector fields
$f_{1,2}^\mu$ can be expressed in terms of the hydrodynamical
currents $j^{\mu}_{f}$ and the polarizations $\lambda^{\mu}_{f}$
of different fermions in matter (see, e.g.,
Refs.~\cite{LobStu01,GriLobStu02,DvoStu02JHEP}),
\begin{equation}\label{deffmu}
  f_{1,2}^\mu=\sqrt{2}G_F\sum\limits_{f=e,p,n}
  \left(
    j^{\mu}_{f}\rho^{(1,2)}_{f}+\lambda^{\mu}_{f}\kappa^{(1,2)}_{f}
  \right),
\end{equation}
where $G_F$ is the Fermi constant and
\begin{align}\label{defq1q2}
  \rho^{(1)}_{f}&=
  (I_{3L}^{(f)}-2Q^{(f)}\sin^{2}\theta_{W}+\delta_{ef}),
  \notag
  \\
  \rho^{(2)}_{f}&=
  (I_{3L}^{(f)}-2Q^{(f)}\sin^{2}\theta_{W}),
  \notag
  \\
  %\quad
  \kappa^{(1)}_{f}&=
  -(I_{3L}^{(f)}+\delta_{ef}),
  %\\
  \quad% \quad \quad \ \
  \kappa^{(2)}_{f}=
  -I_{3L}^{(f)},
  \\
  \delta_{ef}&=
   \begin{cases}
     1, & \text{$f=e$;}
     \\
     0, & \text{$f=n,$ $p$.}
   \end{cases}
  \notag
\end{align}
Here $I_{3L}^{(f)}$ is the third isospin component of the matter
fermion $f$, $Q^{(f)}$ is its electric charge and $\theta_{W}$ is
the Weinberg angle. The hydrodynamical currents and the
polarizations are related to the fermions velocities
$\mathbf{v}_{f}$ and the spin vectors $\bm{\zeta}_{f}$ by means of
the following formulas
\begin{align}\label{jlambda}
  j^{\mu}_{f} & =
  (n_{f},n_{f}\mathbf{v}_{f}),
  \\ \notag
  {\lambda}^{\mu}_{f} & =
  \left(
    n_{f}(\bm{\zeta}_{f}\mathbf{v}_{f}),
    n_{f}\bm{\zeta}_{f}\sqrt{1-v^{2}_{f}}+
    \frac{n_{f}\mathbf{v}_{f}(\bm{\zeta}_{f}\mathbf{v}_{f})}
    {1+\sqrt{1-v^{2}_{f}}}
  \right).
\end{align}
The detailed derivation of Eqs.~\eqref{deffmu}-\eqref{jlambda} is
presented in Refs.~\cite{LobStu01,GriLobStu02,DvoStu02JHEP}.

Following the results of our previous work \cite{Dvo05} we will
study the evolution of the system \eqref{Lagrnu} by solving the
Cauchy problem. Let us choose the initial conditions in the form
\begin{equation}\label{inicond}
  \nu_1(\mathbf{r},0)=0,
  \quad
  \nu_2(\mathbf{r},0)=\xi(\mathbf{r}),
\end{equation}
where $\xi(\mathbf{r})$ is the known function. If one considers
the fermions $\nu_k$ as flavor neutrinos, the initial conditions
in Eq.~\eqref{inicond} correspond to the common situation in a
neutrino oscillations experiment, i.e. $\nu_{\mu,\tau}$ are absent
initially and $\nu_e$ has some known field distribution. We will
be interested in searching for the fields distributions
$\nu_k(\mathbf{r},t)$ for $t>0$.

In order to solve the Cauchy problem we introduce the new set of
the fermions $(\psi_1,\psi_2)$ by means of the matrix
transformation,
\begin{equation}\label{mixmatr}
 \begin{pmatrix}
    \nu_{1} \\
    \nu_{2} \
  \end{pmatrix}=
  \begin{pmatrix}
    \cos \theta & \sin \theta \\
    -\sin \theta & \cos \theta \
  \end{pmatrix}
  \begin{pmatrix}
    \psi_{1} \\
    \psi_{2} \
  \end{pmatrix}.
\end{equation}
The mixing matrix (which is parameterized with help of the one
angle $\theta$) in Eq.~\eqref{mixmatr} is chosen in the way to
eliminate the second and the third terms in Eq.~\eqref{Lagrnu}. If
we had studied the evolution of our system without the external
fields $f_{1,2}^\mu$, i.e. in vacuum, the fermions $\psi_{k}$
would have been called mass eigenstates because they would have
diagonalized the Lagrangian and therefore $\psi_{k}$ would have
had definite masses. In our case ($f_{1,2}^\mu\not=0$), if we
simplify the vacuum mixing terms, the matter term becomes more
complicated compared to Eq.~\eqref{Lagrnu}. The Lagrangian
rewritten in terms of the fields $\psi_k$ can be expressed in the
following way,
\begin{align}
  \mathcal{L}(\psi_1,\psi_2)= & \sum_{k=1,2}\mathcal{L}_0(\psi_k)
  -
  [\bar{\psi}_1\gamma_\mu^L\psi_1(c^2 f_1^{\mu}+s^2 f_2^{\mu})
  \notag \\
  & +
  \bar{\psi}_2\gamma_\mu^L\psi_2(c^2 f_2^{\mu}+s^2 f_1^{\mu})
  \notag
  \\
  \label{Lagrpsi}
  & +
  sc(\bar{\psi}_1\gamma_\mu^L\psi_2+\bar{\psi}_2\gamma_\mu^L\psi_1)
  (f_1^{\mu}-f_2^{\mu})],
\end{align}
where
\begin{equation*}
  \mathcal{L}_0(\psi_k)=
  \bar{\psi}_k(i\gamma^\mu\partial_\mu-m_k)\psi_k.
\end{equation*}
It should be noted that the masses $m_k$ of the fermions $\psi_k$
are related to the masses $\mathfrak{m}_k$ by the formula,
\begin{equation*}
  m_1=c^2\mathfrak{m}_1+s^2\mathfrak{m}_2,
  \quad
  m_2=c^2\mathfrak{m}_2+s^2\mathfrak{m}_1,
\end{equation*}
In Eq.~\eqref{Lagrpsi} we use the notations $c=\cos\theta$ and
$s=\sin\theta$.

Eq.~\eqref{Lagrpsi} has some advantages in comparison with
Eq.~\eqref{Lagrnu} despite of the more complicated matter
interaction term. The terms $g\bar{\nu}_2\nu_1$ and
$g^*\bar{\nu}_1\nu_2$ in Eq.~\eqref{Lagrnu}, which are responsible
for the vacuum oscillations cannot be treated within the
perturbation theory. In order to describe the flavor changing
processes one has to take into account these terms exactly. In
Eq.~\eqref{Lagrpsi} we have the additional interaction terms which
can be analyzed in the common way with help of the perturbation
theory if the strength of the fields $f_{1,2}^\mu$ is supposed to
be weak. The criterion of the external fields weakness is
discussed in details in Sec.~\ref{FIELD}. It is also possible to
assume the convenient time dependence (sometimes it is necessary
to consider the adiabatic switching-on of the interaction) of the
fields $f_{1,2}^\mu$ in Eq.~\eqref{Lagrpsi}. On the contrary we
cannot "switch-on" or "switch-off" the constant $g$ in
Eq.~\eqref{Lagrnu} at some moments of time because it is related
to the properties of a theoretical scheme underlying the
phenomenological model of neutrino oscillations.

\begin{figure}
  \includegraphics[scale=.8]{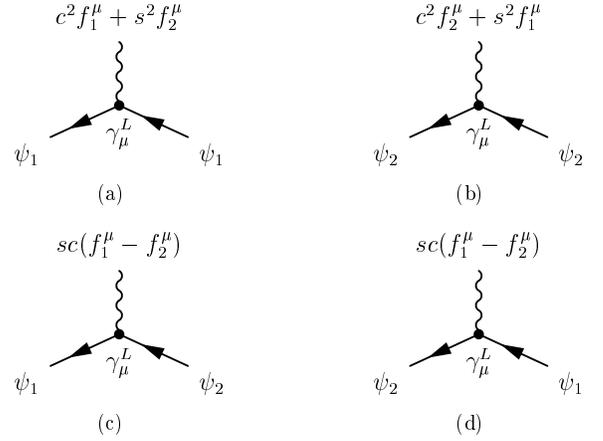}
  \caption{Feynman diagrams contributing to the interaction of
  the fermions $\psi_k$ with the external axial-vector
  fields $f_{1,2}^\mu$.\label{Feyndiag}}
\end{figure}

Now let us discuss the evolution of the system \eqref{Lagrpsi}
with the initial conditions
\begin{equation*}
  \psi_1(\mathbf{r},0)=-s\xi(\mathbf{r}),
  \quad
  \psi_2(\mathbf{r},0)=c\xi(\mathbf{r}),
\end{equation*}
which follow from Eqs.~\eqref{inicond} and \eqref{mixmatr}. We
also suppose that the external fields $f_{1,2}^\mu$ are weak and
one is able to take them into account in the lowest order of the
perturbation theory. If we had solved this problem using the
quantum field theory, we would have calculated the contributions
of the four Feynman diagrams shown on Fig.~\ref{Feyndiag}.

We can always rewrite the Dirac equations which result from
Eq.~\eqref{Lagrpsi} in the form,
\begin{align}\label{psiHamform}
  i\dot{\psi}_1 & =(H_1+V_1)\psi_1+V\psi_2,
  \notag \\
  i\dot{\psi}_2 & =(H_2+V_2)\psi_2+V\psi_1,
\end{align}
where $V_1=\beta\gamma_\mu^L(c^2 f_1^{\mu}+s^2 f_2^{\mu})$,
$V_2=\beta\gamma_\mu^L(c^2 f_2^{\mu}+s^2 f_1^{\mu})$,
$V=sc\beta\gamma_\mu^L(f_1^{\mu}-f_2^{\mu})$ and
$H_k=-i\bm{\alpha}\mathbf{\nabla}+\beta m_k$ are the free fields
Hamiltonians. We are searching for the solutions of
Eqs.~\eqref{psiHamform} in the following way,
\begin{multline}\label{psigen}
  \psi_k(\mathbf{r},t)=
  \\
  \int
  \frac{\mathrm{d}^3\mathbf{p}}{(2\pi)^{3/2}}
  [a_k(\mathbf{p},t)\Psi^{+}_{k,\mathbf{p}}(x)+
  b_k(\mathbf{p},t)\Psi^{-}_{k,\mathbf{p}}(x)],
\end{multline}
where $\Psi^{+}_{k,\mathbf{p}}(x)=u_k(\mathbf{p})e^{-ip_k x}$ and
$\Psi^{-}_{k,\mathbf{p}}(x)=v_k(\mathbf{p})e^{ip_k x}$ are the
basis spinors, $x^\mu=(t,\mathbf{r})$,
$p^\mu_k=(\mathcal{E}_k,\mathbf{p})$ and
$\mathcal{E}_k=\sqrt{\mathbf{p}^2+m_k^2}$ is the energy of the
fermion $\psi_k$. The coefficients $a_k(\mathbf{p},t)$ and
$b_k(\mathbf{p},t)$ are not the creation and annihilation
operators since we are using here the classical field theory. The
values of these functions should be chosen in the way to satisfy
the initial conditions \eqref{inicond}. Note that there is the
additional time dependence of these functions in contrast to the
case of the flavor changing process in vacuum (see
Ref.~\cite{Dvo05}).

Using the orthonormality condition of the basis spinors
$\Psi^{+}_{k,\mathbf{p}}(x)=u_k(\mathbf{p})e^{-ip_k x}$ and
$\Psi^{-}_{k,\mathbf{p}}(x)=v_k(\mathbf{p})e^{ip_k x}$ we obtain
from Eqs.~\eqref{psiHamform} the new integro-differential
equations for the functions $a_k(\mathbf{p},t)$ and
$b_k(\mathbf{p},t)$,
\begin{align*}%\label{abexact}
  i\dot{a}_1(\mathbf{p},t) = & \frac{1}{(2\pi)^{3/2}}
  \bigg(
    \int\mathrm{d}^3\mathbf{r}
    \Psi^{+\dag}_{1,\mathbf{p}}(x) V_1 \psi_1(\mathbf{r},t)
    %\notag
    \\
    & +
    \int\mathrm{d}^3\mathbf{r}
    \Psi^{+\dag}_{1,\mathbf{p}}(x) V \psi_2(\mathbf{r},t)
  \bigg),
  %\displaybreak[1]
  \notag \\
  i\dot{b}_1(\mathbf{p},t) = & \frac{1}{(2\pi)^{3/2}}
  \bigg(
    \int\mathrm{d}^3\mathbf{r}
    \Psi^{-\dag}_{1,\mathbf{p}}(x) V_1 \psi_1(\mathbf{r},t)
    %\notag
    \\
    & +
    \int\mathrm{d}^3\mathbf{r}
    \Psi^{-\dag}_{1,\mathbf{p}}(x) V \psi_2(\mathbf{r},t)
  \bigg),
  \notag \\
  i\dot{a}_2(\mathbf{p},t) = & \frac{1}{(2\pi)^{3/2}}
  \bigg(
    \int\mathrm{d}^3\mathbf{r}
    \Psi^{+\dag}_{2,\mathbf{p}}(x) V_2 \psi_2(\mathbf{r},t)
    \notag \\
    & +
    \int\mathrm{d}^3\mathbf{r}
    \Psi^{+\dag}_{2,\mathbf{p}}(x) V \psi_1(\mathbf{r},t)
  \bigg),
\end{align*}
\begin{align}\label{abexact}
  %\notag \\
  i\dot{b}_2(\mathbf{p},t) = & \frac{1}{(2\pi)^{3/2}}
  \bigg(
    \int\mathrm{d}^3\mathbf{r}
    \Psi^{-\dag}_{2,\mathbf{p}}(x) V_2 \psi_2(\mathbf{r},t)
    \notag \\
    & +
    \int\mathrm{d}^3\mathbf{r}
    \Psi^{-\dag}_{2,\mathbf{p}}(x) V \psi_1(\mathbf{r},t)
  \bigg).
\end{align}
These equations correctly take into account both the Lorentz
invariance and the interaction with the external fields
$f_{1,2}^\mu$. However if we suppose that these external fields
are rather weak, it is possible to look for the solutions of
Eqs.~\eqref{abexact} in the form of the series,
\begin{align}\label{abseries}
  a_k(\mathbf{p},t)= & a^{(0)}_k(\mathbf{p})+
  a^{(1)}_k(\mathbf{p},t)+\dots,
  \notag \\
  b_k(\mathbf{p},t)= & b^{(0)}_k(\mathbf{p})+
  b^{(1)}_k(\mathbf{p},t)+\dots.
\end{align}
Eqs.~\eqref{abseries} mean that the fields distributions can be
also presented in the form of the series $\psi_k(\mathbf{r},t)=
\psi^{(0)}_k(\mathbf{r},t)+\psi^{(1)}_k(\mathbf{r},t)+\dotsb$.
Note that the coefficients $a^{(0)}_k(\mathbf{p})$ and
$b^{(0)}_k(\mathbf{p})$, which correspond to the function
$\psi^{(0)}_k(\mathbf{r},t)$, do not depend on time. The functions
$\psi^{(0)}_k(\mathbf{r},t)$ are responsible for the evolution of
the considered system in vacuum, i.e. at $f_{1,2}^\mu=0$. These
functions in two dimensional space-time have been found in our
previous work \cite{Dvo05} where the analogous Cauchy problem has
been solved in the explicit form in (1+1)-dimensional case.
However to describe the evolution of our system with the non-zero
external fields in the (3+1)-dimensional space-time we should
study the vacuum case also in (3+1) dimensions.

\section{Evolution of the system in vacuum}\label{VACUUM}

To study the behavior of $\psi^{(0)}_k(\mathbf{r},t)$, i.e. the
evolution of our system in vacuum, we use the results of
Ref.~\cite{Dvo05} where it was revealed that the fields
distributions for the given initial conditions have the form,
\begin{align}\label{psiSr}
  \psi^{(0)}_1(\mathbf{r},t) = &
  -s\int\mathrm{d}^3\mathbf{r}'
  S_1(\mathbf{r}'-\mathbf{r},t)(-i\beta)\xi(\mathbf{r}'),
  \notag \\
  \psi^{(0)}_2(\mathbf{r},t) = &
  c\int\mathrm{d}^3\mathbf{r}'
  S_2(\mathbf{r}'-\mathbf{r},t)(-i\beta)\xi(\mathbf{r}'),
\end{align}
where $S_k(\mathbf{r},t)$ is the Pauli-Jordan function for the
fermion $\psi_k$ (see, e.g., Ref.~\cite{BogShi80p607full}). It
should be noted that Eqs.~\eqref{psiSr} are the most general ones
and valid in (3+1) dimensions. Contrary to the approach of
Ref.~\cite{Dvo05} here we use the momentum representation because
the integrations in Eqs.~\eqref{psiSr} are rather cumbersome in
the coordinate representation. Thus one rewrites these expressions
using the Fourier transform of the initial conditions.
Eqs.~\eqref{psiSr} now take the form,
\begin{align}\label{psi0p}
  \psi^{(0)}_1(\mathbf{r},t) & =
  -s\int\frac{\mathrm{d}^3\mathbf{p}}{(2\pi)^3}
  e^{i\mathbf{p}\mathbf{r}}
  S_1(-\mathbf{p},t)(-i\beta)\xi(\mathbf{p}),
  \notag \\
  \psi^{(0)}_2(\mathbf{r},t) & =
  c\int\frac{\mathrm{d}^3\mathbf{p}}{(2\pi)^3}
  e^{i\mathbf{p}\mathbf{r}}
  S_2(-\mathbf{p},t)(-i\beta)\xi(\mathbf{p}),
\end{align}
where
\begin{equation}\label{PJp}
  S_k(-\mathbf{p},t)=
  \left[
    \cos\mathcal{E}_k t-
    i\frac{\sin\mathcal{E}_k t}{\mathcal{E}_k}
    (\bm{\alpha}\mathbf{p}+\beta m_k)
  \right](i\beta),
\end{equation}
and
\begin{equation*}
  \xi(\mathbf{p})=\int\mathrm{d}^3\mathbf{p}
  e^{-i\mathbf{p}\mathbf{r}}\xi(\mathbf{r}),
\end{equation*}
are the Fourier transforms of the functions $S_k(\mathbf{r},t)$
and $\xi(\mathbf{r})$.

Now let us choose the initial condition. We suppose that the
initial field distribution of $\nu_2$ is the plane wave, i.e.
$\xi(\mathbf{r})=e^{i\bm{\omega}\mathbf{r}}\xi_0$, where $\xi_0$
is the normalization spinor. The Fourier transform of this
function can be simply computed,
$\xi(\mathbf{p})=(2\pi)^3\delta^3(\bm{\omega}-\mathbf{p})\xi_0$.
Using Eqs.~\eqref{psi0p} we get the fields distributions in the
(3+1)-dimensional space-time for the plane wave initial condition,
\begin{align}\label{psi0pw}
  \psi^{(0)}_1(\mathbf{r},t) = &
  -se^{i\bm{\omega}\mathbf{r}}
  \bigg[
    \cos[\mathcal{E}_1(\omega) t]
    \notag \\
    & -
    i\frac{\sin[\mathcal{E}_1(\omega) t]}{\mathcal{E}_1(\omega)}
    (\bm{\alpha}\bm{\omega}+\beta m_1)
  \bigg]\xi_0,
  \notag \\
  \psi^{(0)}_2(\mathbf{r},t) = &
  ce^{i\bm{\omega}\mathbf{r}}
  \bigg[
    \cos[\mathcal{E}_2(\omega) t]
    \notag \\
    & -
    i\frac{\sin[\mathcal{E}_2(\omega) t]}{\mathcal{E}_2(\omega)}
    (\bm{\alpha}\bm{\omega}+\beta m_2)
  \bigg]\xi_0,
\end{align}
where $\mathcal{E}_k(\omega)=\sqrt{\omega^2+m_k^2}$.

In the following we discuss the case of the rapidly oscillating
initial conditions, i.e. $\omega\gg m_{1,2}$. One obtains from
Eqs.~\eqref{psi0pw} the fields distributions for $\omega\gg
m_{1,2}$ in the following form,
\begin{align}\label{psi0hf}
  \psi^{(0)}_1(\mathbf{r},t) = &
  -se^{i\bm{\omega}\mathbf{r}}
  (\cos[\mathcal{E}_1(\omega) t]
  \notag \\
  & -
  i(\bm{\alpha}\mathbf{n})
  \sin[\mathcal{E}_1(\omega) t])\xi_0,
  \notag \\
  \psi^{(0)}_2(\mathbf{r},t) = &
  ce^{i\bm{\omega}\mathbf{r}}
  (\cos[\mathcal{E}_2(\omega) t]
  \notag \\
  & -
  i(\bm{\alpha}\mathbf{n})
  \sin[\mathcal{E}_2(\omega) t])\xi_0,
\end{align}
where $\mathbf{n}=\bm{\omega}/\omega$ is the unit vector in the
direction of the initial field distribution momentum. The fermion
$\nu_1$ is absent at $t=0$. Therefore it would be interesting to
examine the field distribution $\nu^{(0)}_1(\mathbf{r},t)$ at
$t>0$. Using Eqs.~\eqref{mixmatr} and \eqref{psi0hf} we obtain,
\begin{align}\label{nu01}
  \nu^{(0)}_1(\mathbf{r},t)= &
  c\psi^{(0)}_1+s\psi^{(0)}_2=
  \sin2\theta\sin[\Delta(\omega)t]
  \\
  & \times
  \notag
  \{\sin[\sigma(\omega)t]+
  i(\bm{\alpha}\mathbf{n})\cos[\sigma(\omega)t]\}
  e^{i\bm{\omega}\mathbf{r}}\xi_0,
\end{align}
where
\begin{align*}
  \sigma(\omega) & = \frac{\mathcal{E}_1(\omega)+\mathcal{E}_2(\omega)}{2}
  \to\omega+\frac{m^2_1+m^2_2}{4\omega},
  \\
  \Delta(\omega) & = \frac{\mathcal{E}_1(\omega)-\mathcal{E}_2(\omega)}{2}
  \to\frac{m^2_1-m^2_2}{4\omega}=
  \frac{\Delta m^2}{4\omega}.
\end{align*}
The measurable quantity of a classical spinor field is the
intensity. With help of Eq.~\eqref{psi0hf} one gets the intensity
of the fermion $\nu^{(0)}_1$ in the following form,
\begin{align}\label{I01}
  I^{(0)}_1(t)= & |\nu^{(0)}_1(\mathbf{r},t)|^2=
  \sin^2(2\theta)\sin^2[\Delta(\omega) t]
  \notag \\
  & \times
  \xi^\dag_0
  |\sin[\sigma(\omega) t]+i(\bm{\alpha}\mathbf{n})
  \cos[\sigma(\omega) t]|^2
  \xi_0
  \notag \\
  & = \sin^2(2\theta)
  \sin^2
  \left(
    \frac{\Delta m^2}{4\omega} t
  \right),
\end{align}
which reproduces the Pontecorvo formula in the (3+1)-dimensional
space-time since we can regard the intensity of the fermion
$\nu_1$ as the transition probability in two neutrino system.
Eq.~\eqref{I01} also generalizes the result of our previous work
\cite{Dvo05} where the analogous expression was derived in (1+1)
dimensions.

\begin{widetext}

\section{Interaction of the system with an external field}
\label{FIELD}

In order to proceed in our studying of the two neutrino system
evolution under the influence of the external fields $f_{1,2}^\mu$
we discuss the further correction to the vacuum case. The first
order corrections to Eqs.~\eqref{psi0p} can be derived from
Eqs.~\eqref{psigen} and \eqref{abexact} and have the form,
\begin{align}\label{psi1S}
  \psi^{(1)}_1(\mathbf{r},t) = &
  i\int\frac{\mathrm{d}^3\mathbf{p}}{(2\pi)^3}
  \frac{e^{i\mathbf{p}\mathbf{r}}}{2\mathcal{E}_1}
  \{\mathcal{E}_1
  [s V_1(\mathfrak{S}^{+}_{1}e^{-i\mathcal{E}_1 t}+
  \mathfrak{S}^{-}_{1}e^{+i\mathcal{E}_1 t})
  %\notag \\
  %&
  -
  c V(\mathfrak{S}^{+}_{12}e^{-i\mathcal{E}_1 t}+
  \mathfrak{S}^{-}_{12}e^{+i\mathcal{E}_1 t})]
  \notag \\ & +
  (\bm{\alpha}\mathbf{p}+\beta m_1)
  [s V_1(\mathfrak{S}^{+}_{1}e^{-i\mathcal{E}_1 t}-
  \mathfrak{S}^{-}_{1}e^{+i\mathcal{E}_1 t})
  %\notag \\
  %&
  -
  c V(\mathfrak{S}^{+}_{12}e^{-i\mathcal{E}_1 t}-
  \mathfrak{S}^{-}_{12}e^{+i\mathcal{E}_1 t})]\}
  (-i\beta)\xi(\mathbf{p}),
  \notag \\
  \psi^{(1)}_2(\mathbf{r},t) = &
  -i\int\frac{\mathrm{d}^3\mathbf{p}}{(2\pi)^3}
  \frac{e^{i\mathbf{p}\mathbf{r}}}{2\mathcal{E}_2}
  \{\mathcal{E}_2
  [c V_2(\mathfrak{S}^{+}_{2}e^{-i\mathcal{E}_2 t}+
  \mathfrak{S}^{-}_{2}e^{+i\mathcal{E}_2 t})
  %\notag \\
  %&
  -
  s V(\mathfrak{S}^{+}_{21}e^{-i\mathcal{E}_2 t}+
  \mathfrak{S}^{-}_{21}e^{+i\mathcal{E}_2 t})]
  \notag \\ & +
  (\bm{\alpha}\mathbf{p}+\beta m_2)
  [c V_2(\mathfrak{S}^{+}_{2}e^{-i\mathcal{E}_2 t}-
  \mathfrak{S}^{-}_{2}e^{+i\mathcal{E}_2 t})
  %\notag \\
  %&
  -
  s V(\mathfrak{S}^{+}_{21}e^{-i\mathcal{E}_2 t}-
  \mathfrak{S}^{-}_{21}e^{+i\mathcal{E}_2 t})]\}
  (-i\beta)\xi(\mathbf{p}),
\end{align}
where
\begin{equation}\label{Sfun}
  \mathfrak{S}^{\pm}_{12} =
  \int_0^t
  e^{\pm i\mathcal{E}_1 t}S_2(-\mathbf{p},t)\mathrm{d}t,
  \quad
  %\notag \\
  \mathfrak{S}^{\pm}_{21} =
  \int_0^t
  e^{\pm i\mathcal{E}_2 t}S_1(-\mathbf{p},t)\mathrm{d}t,
  \quad
  %\notag \\
  \mathfrak{S}^{\pm}_{1,2} =
  \int_0^t
  e^{\pm i\mathcal{E}_{1,2} t}S_{1,2}(-\mathbf{p},t)\mathrm{d}t.
\end{equation}
When we derive Eqs.~\eqref{psi1S} we suppose that
$a^{(1)}_{1,2}(\mathbf{p},0)=b^{(1)}_{1,2}(\mathbf{p},0)=0$. It
means that at $t=0$ the fields distributions are determined by
Eqs.~\eqref{psi0p}. One can find the explicit form of the
functions given in Eqs.~\eqref{Sfun} using Eq.~\eqref{PJp},
%
%\begin{widetext}
\begin{align}\label{Sexplicit}
  \mathfrak{S}^{\pm}_{12} & =
  \frac{1}{2}
  \left\{
    e^{\pm i\sigma t}\frac{\sin\sigma t}{\sigma}+
    e^{\pm i\Delta t}\frac{\sin\Delta t}{\Delta}-
    \frac{\bm{\alpha}\mathbf{p}+\beta m_2}{\mathcal{E}_2}
    \left(
      \pm e^{\pm i\sigma t}\frac{\sin\sigma t}{\sigma}
      \mp e^{\pm i\Delta t}\frac{\sin\Delta t}{\Delta}
    \right)
  \right\}(i\beta),
  \notag \\
  \mathfrak{S}^{\pm}_{21} & =
  \frac{1}{2}
  \left\{
    e^{\pm i\sigma t}\frac{\sin\sigma t}{\sigma}+
    e^{\mp i\Delta t}\frac{\sin\Delta t}{\Delta}-
    \frac{\bm{\alpha}\mathbf{p}+\beta m_1}{\mathcal{E}_1}
    \left(
      \pm e^{\pm i\sigma t}\frac{\sin\sigma t}{\sigma}
      \mp e^{\mp i\Delta t}\frac{\sin\Delta t}{\Delta}
    \right)
  \right\}(i\beta),
  \notag \\
  \mathfrak{S}^{\pm}_{1,2} & =
  \frac{1}{2}
  \left\{
    \left(
      \pm\frac{1}{2i\mathcal{E}_{1,2}}
      e^{\pm 2i\mathcal{E}_{1,2}t}+t
    \right)+
    i\frac{\bm{\alpha}\mathbf{p}+\beta m_{1,2}}{\mathcal{E}_{1,2}}
    \left(
      \frac{1}{2i\mathcal{E}_{1,2}}
      e^{\pm 2i\mathcal{E}_{1,2}t}\mp t
    \right)
  \right\}(i\beta),
\end{align}
where
\begin{equation*}
  \sigma=\frac{\mathcal{E}_1+\mathcal{E}_2}{2},
  \quad
  \Delta=\frac{\mathcal{E}_1-\mathcal{E}_2}{2}.
\end{equation*}
On the basis of Eqs.~\eqref{psi1S} and \eqref{Sexplicit} we obtain
the expressions for the first order corrections to the vacuum case
in the following form,
\begin{align}\label{psi1p}
  \psi^{(1)}_1(\mathbf{r},t) = &
  i\int\frac{\mathrm{d}^3\mathbf{p}}{(2\pi)^3}
  \frac{e^{i\mathbf{p}\mathbf{r}}}{2\mathcal{E}_1}
  \bigg\{
    \mathcal{E}_1
    \bigg[
      s V_1
      \left\{
        \frac{\sin\mathcal{E}_1 t}{\mathcal{E}_1}+
        t\cos\mathcal{E}_1 t-
        i\frac{\bm{\alpha}\mathbf{p}+\beta m_1}{\mathcal{E}_1}
        t\sin\mathcal{E}_1 t
      \right\}
      \notag \\
      \displaybreak[1]
      & -
      c V
      \left\{
        \frac{\sin\sigma t}{\sigma}\cos\Delta t+
        \frac{\sin\Delta t}{\Delta}\cos\sigma t+
        i\frac{\bm{\alpha}\mathbf{p}+\beta m_2}{\mathcal{E}_2}
        \sin\sigma t\sin\Delta t
        \left(
          \frac{1}{\sigma}-\frac{1}{\Delta}
        \right)
      \right\}
    \bigg]
    \notag \\ & +
    (\bm{\alpha}\mathbf{p}+\beta m_1)
    \bigg[
      s V_1
      \left\{
        -it\sin\mathcal{E}_1 t+
        \frac{\bm{\alpha}\mathbf{p}+\beta m_1}{\mathcal{E}_1}
        \left(
          t\cos\mathcal{E}_1 t-
          \frac{\sin\mathcal{E}_1 t}{\mathcal{E}_1}
        \right)
      \right\}
      \notag \\
      &
      -
      c V
      \bigg\{
        -i\sin\sigma t\sin\Delta t
        \left(
          \frac{1}{\sigma}+\frac{1}{\Delta}
        \right)
        %\notag \\
        %&
        +
        \frac{\bm{\alpha}\mathbf{p}+\beta m_2}{\mathcal{E}_2}
        \left(
          \frac{\sin\Delta t}{\Delta}\cos\sigma t-
          \frac{\sin\sigma t}{\sigma}\cos\Delta t
        \right)
      \bigg\}
    \bigg]
  \bigg\}
  \xi(\mathbf{p}),
  \notag \\
  \displaybreak[1]
  \psi^{(1)}_2(\mathbf{r},t) = &
  -i\int\frac{\mathrm{d}^3\mathbf{p}}{(2\pi)^3}
  \frac{e^{i\mathbf{p}\mathbf{r}}}{2\mathcal{E}_2}
  \bigg\{
    \mathcal{E}_2
    \bigg[
      c V_2
      \left\{
        \frac{\sin\mathcal{E}_2 t}{\mathcal{E}_2}+
        t\cos\mathcal{E}_2 t-
        i\frac{\bm{\alpha}\mathbf{p}+\beta m_2}{\mathcal{E}_2}
        t\sin\mathcal{E}_2 t
      \right\}
      \notag \\
      \displaybreak[1]
      & -
      s V
      \left\{
        \frac{\sin\sigma t}{\sigma}\cos\Delta t+
        \frac{\sin\Delta t}{\Delta}\cos\sigma t-
        i\frac{\bm{\alpha}\mathbf{p}+\beta m_1}{\mathcal{E}_1}
        \sin\sigma t\sin\Delta t
        \left(
          \frac{1}{\sigma}+\frac{1}{\Delta}
        \right)
      \right\}
    \bigg]
    \notag \\
    \displaybreak[1]
    & +
    (\bm{\alpha}\mathbf{p}+\beta m_2)
    \bigg[
      c V_2
      \left\{
        -it\sin\mathcal{E}_2 t+
        \frac{\bm{\alpha}\mathbf{p}+\beta m_2}{\mathcal{E}_2}
        \left(
          t\cos\mathcal{E}_2 t-
          \frac{\sin\mathcal{E}_2 t}{\mathcal{E}_2}
        \right)
      \right\}
      \notag \\
      \displaybreak[1]
      & -
      s V
      \bigg\{
        i\sin\sigma t\sin\Delta t
        \left(
          \frac{1}{\sigma}-\frac{1}{\Delta}
        \right)
        %\notag \\
        %&
        +
        \frac{\bm{\alpha}\mathbf{p}+\beta m_1}{\mathcal{E}_1}
        \left(
          \frac{\sin\Delta t}{\Delta}\cos\sigma t-
          \frac{\sin\sigma t}{\sigma}\cos\Delta t
        \right)
      \bigg\}
    \bigg]
  \bigg\}
  \xi(\mathbf{p}).
\end{align}
%\end{widetext}
%
Note that these expressions are valid for the arbitrary initial
conditions $\xi(\mathbf{p})$ and exactly take into account the
Lorentz invariance. It should be also mentioned that along with
the harmonic functions there are several terms in the integrands
which linearly depend on time. Therefore for Eqs.~\eqref{psi1p} to
be meaning-bearing one has to assume that the potentials $V_{1,2}$
and $V$ are rather weak, i.e. we study the interaction of our
system with weak external fields (see Eq.~\eqref{weak} below).

The integrations in Eqs.~\eqref{psi1p} are rather complicated for
the arbitrary initial conditions. That is why we choose the
function $\xi(\mathbf{p})$ analogously to Sec.~\ref{VACUUM}, i.e.
we again suppose that
$\xi(\mathbf{p})=(2\pi)^3\delta^3(\bm{\omega}-\mathbf{p})\xi_0$.
The integrations over momenta are eliminated with help of the
$\delta$-functions. We also consider high frequency approximation,
i.e. $\omega\gg m_{1,2}$. Finally we obtain the following
expressions for the fields distributions of the fermions $\psi_k$,
\begin{align}\label{psi1hf}
  \psi^{(1)}_1(\mathbf{r},t) = &
  e^{i\bm{\omega}\mathbf{r}}
  \frac{i}{2}
  \bigg\{
    st[V_1+(\bm{\alpha}\mathbf{n})V_1(\bm{\alpha}\mathbf{n})]
    (\cos[\mathcal{E}_1(\omega)t]-
    i(\bm{\alpha}\mathbf{n})\sin[\mathcal{E}_1(\omega)t])
    \notag
    \displaybreak[1]
    \\
    & -
    c\frac{\sin[\Delta(\omega)t]}{\Delta(\omega)}
    [V+(\bm{\alpha}\mathbf{n})V(\bm{\alpha}\mathbf{n})]
    (\cos[\sigma(\omega)t]-
    i(\bm{\alpha}\mathbf{n})\sin[\sigma(\omega)t])
  \bigg\}\xi_0,
  \notag
  \displaybreak[1]
  \\
  \psi^{(2)}_1(\mathbf{r},t) = &
  -e^{i\bm{\omega}\mathbf{r}}
  \frac{i}{2}
  \bigg\{
    ct[V_2+(\bm{\alpha}\mathbf{n})V_2(\bm{\alpha}\mathbf{n})]
    (\cos[\mathcal{E}_2(\omega)t]-
    i(\bm{\alpha}\mathbf{n})\sin[\mathcal{E}_2(\omega)t])
    \notag \\
    & -
    s\frac{\sin[\Delta(\omega)t]}{\Delta(\omega)}
    [V+(\bm{\alpha}\mathbf{n})V(\bm{\alpha}\mathbf{n})]
    (\cos[\sigma(\omega)t]-
    i(\bm{\alpha}\mathbf{n})\sin[\sigma(\omega)t])
  \bigg\}\xi_0.
\end{align}
%\end{widetext}
%
On the basis of Eqs.~\eqref{mixmatr} and \eqref{psi1hf} we can
derive the first order correction to the field distribution of the
fermion $\nu_1$
%$\nu^{(1)}_1=c\psi^{(1)}_1+s\psi^{(1)}_2$
in the form,
\begin{align}\label{nu11}
  \nu^{(1)}_1(\mathbf{r},t)= &
  -\sin2\theta e^{i\bm{\omega}\mathbf{r}}
  \frac{i}{4}
  %[f^0-(\mathbf{f}\mathbf{n})(\bm{\Sigma}\mathbf{n})]
  %(1+\gamma^5)
  \bigg\{
    \cos2\theta\frac{\sin\Delta(\omega)t}{\Delta(\omega)}
    (F_1-F_2)
    (\cos[\sigma(\omega)t]-
    i(\bm{\alpha}\mathbf{n})\sin[\sigma(\omega)t])
    \notag
    \\
    & -
    t
    \big[
      F_1
      \{c^2(\cos[\mathcal{E}_1(\omega)t]-
      i(\bm{\alpha}\mathbf{n})\sin[\mathcal{E}_1(\omega)t])
      -
      s^2(\cos[\mathcal{E}_2(\omega)t]-
      i(\bm{\alpha}\mathbf{n})\sin[\mathcal{E}_2(\omega)t])\}
      \notag
      \\
      & -
      F_2
      \{c^2(\cos[\mathcal{E}_2(\omega)t]-
      i(\bm{\alpha}\mathbf{n})\sin[\mathcal{E}_2(\omega)t])
      -
      s^2(\cos[\mathcal{E}_1(\omega)t]-
      i(\bm{\alpha}\mathbf{n})\sin[\mathcal{E}_1(\omega)t])\}
    \big]
  \bigg\}\xi_0,
\end{align}
\end{widetext}
where $F_{1,2}=
[f_{1,2}^0-(\mathbf{f}_{1,2}\mathbf{n})(\bm{\Sigma}\mathbf{n})]
(1+\gamma^5)$ and $\bm{\Sigma}=-\gamma^5\bm{\alpha}$.

To calculate the intensity of the field $\nu_1$ one should take
into account that the final expression for the intensity must
contain only terms linear in external fields. Therefore the first
order correction to the intensity should be calculated with help
of the formula
\begin{equation*}
  I^{(1)}_1(t)=
  \nu^{(0)\dag}_1\nu^{(1)}_1+\nu^{(1)\dag}_1\nu^{(0)}_1.
\end{equation*}
Using Eqs.~\eqref{nu01} and \eqref{nu11} we get the expression for
$I^{(1)}_1$,
\begin{align}\label{I11}
  I^{(1)}_1(t) & = \sin^2(2\theta)\cos2\theta\sin[\Delta(\omega)t]
  \\
  & \times
  \frac{1}{2}
  \left(
    \frac{\sin[\Delta(\omega)t]}{\Delta(\omega)}-
    t\cos[\Delta(\omega)t]
  \right)
  \notag \\
  & \times
  \langle
    ([f_2^0(\bm{\alpha}\mathbf{n})-(\mathbf{f}_2\mathbf{n})]-
    [f_1^0(\bm{\alpha}\mathbf{n})-(\mathbf{f}_1\mathbf{n})])
    (1+\gamma^5)
  \rangle.
  \notag
\end{align}
In Eq.~\eqref{I11} we use the notation
$\langle(\dots)\rangle=\xi^{\dag}_0(\dots)\xi_0$. To compute the
mean value with help of the normalization spinor $\xi_0$ we can
suppose that
$\xi(\mathbf{r})=\exp(-iE_{\nu_2}t)\xi(\mathbf{r})|_{t\to 0}$.
Then we notice that for spinors corresponding to high energies one
has the obvious identities $(1+\gamma^5)\xi_0\approx 2\xi_0$ and
$\xi^{\dag}_0(\bm{\alpha}\mathbf{n})\xi_0\approx 1$. Putting
together Eqs.~\eqref{I01} and \eqref{I11} we obtain the final
expression for the intensity of the fermion $\nu_1$,
\begin{align}\label{I101}
  I_1(t)= & I^{(0)}_1(t)+I^{(1)}_1(t)
  =
  \notag
  \\ &
  \sin^2(2\theta)
  \bigg\{
    \sin^2[\Delta(\omega)t]+
    \cos2\theta\sin[\Delta(\omega)t]
    \notag \\
    & \times
    \left(
      \frac{\sin[\Delta(\omega)t]}{\Delta(\omega)}-
      t\cos[\Delta(\omega)t]
    \right)
    \notag \\
    & \times
    ([f_2^0-(\mathbf{f}_2\mathbf{n})]-
    [f_1^0-(\mathbf{f}_1\mathbf{n})])
  \bigg\}.
\end{align}
Using Eq.~\eqref{I101} it is possible to define the scope of the
applied method, i.e. we can evaluate the strength of external
fields necessary for the perturbative approach to be valid. With
help of Eq.~\eqref{I101} one obtains the inequalities,
\begin{equation}\label{weak}
  A\cos2\theta\ll \Delta(\omega),
  \quad
  At\cos2\theta\ll 1,
\end{equation}
where
$A=[f_2^0-(\mathbf{f}_2\mathbf{n})]-[f_1^0-(\mathbf{f}_1\mathbf{n})]$.
If Eq.~\eqref{weak} is satisfied, the contribution of external
axial-vector fields to neutrino flavor oscillations is small
compared to the vacuum term. It should be noted that
Eq.~~\eqref{weak} is valid in the ultrarelativistic case. For
neutrinos with $\mathcal{E}_k(\omega) \sim m_k$ we should rely on
Eqs.~\eqref{psi1p} rather than on Eqs.~\eqref{psi1hf}. In this
case the condition of our method applicability will be different
from Eq.~\eqref{weak}.

Now let us compare Eq.~\eqref{I101} with the neutrino transition
probability formula. Flavor neutrinos are considered to interact
with external axial-vector fields as it is described in
Eqs.~\eqref{deffmu}-\eqref{jlambda}. Then the probability to find
muon or $\tau$-neutrinos in the electron neutrinos beam in
presence of moving and polarized matter is expressed in the
following way (see, e.g., Refs.~\cite{NunSemSmiVal97,GriLobStu02})
\begin{equation}\label{etomutau}
  P_{\nu_e\to\nu_{\mu,\tau}}(t)=
  \sin^2(2\theta_\mathrm{eff})\sin^2
  \left(
    \frac{\pi t}{L_\mathrm{eff}}
  \right),
\end{equation}
where
\begin{multline}\label{thetaeff}
  \sin^2(2\theta_\mathrm{eff})=
  \\
  \frac{\Delta^2(\omega)\sin^2(2\theta)}
  {[\Delta(\omega)\cos2\theta-A/2]^2+
  \Delta^2(\omega)\sin^2(2\theta)},
\end{multline}
is the definition of the effective mixing angle and
\begin{equation}\label{Leff}
  \frac{\pi}{L_\mathrm{eff}}=
  \sqrt{[\Delta(\omega)\cos2\theta-A/2]^2+
  \Delta^2(\omega)\sin^2(2\theta)},
\end{equation}
is the definition of the effective oscillation length.

%In Eqs.~\eqref{thetaeff} and \eqref{Leff} we use the notation
%$A=[f_2^0-(\mathbf{f}_2\mathbf{n})]-[f_1^0-(\mathbf{f}_1\mathbf{n})]$.

To discuss the weak external field limit in
Eqs.~\eqref{etomutau}-\eqref{Leff} we should expand
Eqs.~\eqref{thetaeff} and \eqref{Leff} over small parameter $A$
[see Eq.~\eqref{weak}]. As a result one gets
\begin{align}\label{thetaLeffexp}
  \sin^2(2\theta_\mathrm{eff})\approx &
  \sin^2(2\theta)
  \left(
    1+\frac{A\cos2\theta}{\Delta(\omega)}
  \right),
  \notag
  \\
  \pi/L_\mathrm{eff}\approx &
  \Delta(\omega)-(A\cos2\theta)/2.
\end{align}
It is also necessary to expand the time dependent factor in
Eq.~\eqref{etomutau},
\begin{align}\label{phaseexp}
  \sin^2
  \left(
    \frac{\pi t}{L_\mathrm{eff}}
  \right)
  \approx &
  \sin^2[2\Delta(\omega)t]
  \notag \\
  & -
  At\cos2\theta\sin[\Delta(\omega)t]\cos[\Delta(\omega)t].
\end{align}
Note that Eq.~\eqref{phaseexp} is valid while $At\cos2\theta\ll
1$, that coincides with the second inequality in Eq.~\eqref{weak}.
With help of Eqs.~\eqref{thetaLeffexp} and \eqref{phaseexp} the
neutrino transition probability is reduced to the form,
\begin{align*}%\label{Pontorig}
  P_{\nu_e\to\nu_{\mu,\tau}}(t)= & \sin^2(2\theta)
  \bigg\{
    \sin^2[\Delta(\omega)t]+
    \cos2\theta\sin[\Delta(\omega)t]
    \notag \\
    & \times
    \left(
      \frac{\sin[\Delta(\omega)t]}{\Delta(\omega)}-
      t\cos[\Delta(\omega)t]
    \right)
    \notag \\
    & \times
    ([f_2^0-(\mathbf{f}_2\mathbf{n})]-
    [f_1^0-(\mathbf{f}_1\mathbf{n})])
  \bigg\},
\end{align*}
which coincides with Eq.~\eqref{I101}. This comparison shows that
neutrino flavor oscillations in weak axial-vector fields (e.g., if
a neutrino propagates in moving and polarized matter) can be
treated with help of the classical field theory approach.

\section{Conclusion}\label{CONCL}

In conclusion we mention that the evolution of coupled classical
fermions under the influence of external axial-vector fields has
been studied in the present paper. We have discussed the
particular case of two coupled fermions and formulated the Cauchy
problem for this system. If the initial conditions were chosen in
the appropriate way, as it has been shown in Sec.~\ref{GF}, the
described system might serve as a theoretical model of neutrino
flavor oscillations in matter. The initial conditions problem has
been solved with help of the perturbation theory. We have found
the zero and the first order terms in the fields distributions
expansions over the external fields strength. It should be noted
that the obtained results exactly take into account the Lorentz
invariance as well as they are valid in (3+1)-dimensional
space-time. The intensity of the zero order term corresponds to
the case of the neutrino flavor oscillations in vacuum. Therefore
we have generalized our previous calculations performed in (1+1)
dimensions in Ref.~\cite{Dvo05}. The first order correction is
responsible for the neutrino interaction with moving and polarized
matter. We have obtained this intensity of the fermion field
%and shown that it coincides with the neutrino transition
%probability
at great oscillations frequencies of the initial field
distribution, that corresponds to ultrarelativistic neutrinos.
Note that we have compared our results with the transition
probability formula for neutrino flavor oscillation in moving and
polarized matter derived in
Refs.~\cite{NunSemSmiVal97,GriLobStu02} and revealed an agreement
in the case of weak external fields. This comparison proves the
validity of the method elaborated in our work. Finally it has been
demonstrated that neutrino flavor oscillations in moving and
polarized matter could be described with help of the classical
field theory.

It is interesting to notice that along with the usual neutrino
oscillations phase equal to $\Delta m^2/(4\omega)$ the classical
field theory approach also reveals rapid harmonic oscillations on
the frequency [see, e.g., Eqs.~\eqref{psi0pw} and \eqref{psi1p}]
\begin{equation*}
  \omega_\mathrm{rapid}=
  \frac{\mathcal{E}_1(\omega)+\mathcal{E}_2(\omega)}{2}\to
  \omega+
  \frac{m_1^2+m_2^2}{4\omega}.
\end{equation*}
However these terms are suppressed by the ratios $m_k/\omega$
which are small for great values of $\omega$. This case
corresponds to ultrarelativistic neutrinos. The analogous terms
were discussed in many previous publications devoted to neutrino
flavor oscillations (see, e.g., Refs.~\cite{BlaVit95,Beu02}). In
our work it has been demonstrated that such contributions to the
neutrino transition probability appear even in the classical field
theory approach. These terms arise from the accurate account of
the Lorentz invariance.

\begin{acknowledgments}
This research was supported by grant of Russian Science Support
Foundation. The author is indebted to Alexander Grigoriev and
Alexander Studenikin (MSU) for helpful discussions as well as to
Takuya Morozumi for the invitation to visit Hiroshima University.
\end{acknowledgments}

\bibliography{generaleng}

\begin{thebibliography}{21}
\expandafter\ifx\csname natexlab\endcsname\relax\def\natexlab#1{#1}\fi
\expandafter\ifx\csname bibnamefont\endcsname\relax
  \def\bibnamefont#1{#1}\fi
\expandafter\ifx\csname bibfnamefont\endcsname\relax
  \def\bibfnamefont#1{#1}\fi
\expandafter\ifx\csname citenamefont\endcsname\relax
  \def\citenamefont#1{#1}\fi
\expandafter\ifx\csname url\endcsname\relax
  \def\url#1{\texttt{#1}}\fi
\expandafter\ifx\csname urlprefix\endcsname\relax\def\urlprefix{URL }\fi
\providecommand{\bibinfo}[2]{#2}
\providecommand{\eprint}[2][]{\url{#2}}

\bibitem[{\citenamefont{Aharmim et~al.}(2004)}]{Aha04}
\bibinfo{author}{\bibfnamefont{B.}~\bibnamefont{Aharmim}} \bibnamefont{et~al.},
  \bibinfo{journal}{Phys. Rev. D} \textbf{\bibinfo{volume}{70}},
  \bibinfo{pages}{093014} (\bibinfo{year}{2004}).

\bibitem[{\citenamefont{Hosaka et~al.}()}]{Hos05}
\bibinfo{author}{\bibfnamefont{J.}~\bibnamefont{Hosaka}} \bibnamefont{et~al.},
  \emph{\bibinfo{title}{Solar neutrino measurements in
  {S}uper-{K}amiokande-{I}}}, \eprint{hep-ex/0508053}.

\bibitem[{\citenamefont{Kayser}(1981)}]{Kay81}
\bibinfo{author}{\bibfnamefont{B.}~\bibnamefont{Kayser}},
  \bibinfo{journal}{Phys. Rev. D} \textbf{\bibinfo{volume}{24}},
  \bibinfo{pages}{110} (\bibinfo{year}{1981}).

\bibitem[{\citenamefont{Pontecorvo}(1958)}]{Pon58eng}
\bibinfo{author}{\bibfnamefont{B.}~\bibnamefont{Pontecorvo}},
  \bibinfo{journal}{JETP} \textbf{\bibinfo{volume}{7}}, \bibinfo{pages}{172}
  (\bibinfo{year}{1958}).

\bibitem[{\citenamefont{Giunti et~al.}(1993)\citenamefont{Giunti, Kim, Lee, and
  Lee}}]{GiuKimLeeLee93}
\bibinfo{author}{\bibfnamefont{C.}~\bibnamefont{Giunti}},
  \bibinfo{author}{\bibfnamefont{C.~W.} \bibnamefont{Kim}},
  \bibinfo{author}{\bibfnamefont{J.~A.} \bibnamefont{Lee}}, \bibnamefont{and}
  \bibinfo{author}{\bibfnamefont{U.~W.} \bibnamefont{Lee}},
  \bibinfo{journal}{Phys. Rev. D} \textbf{\bibinfo{volume}{48}},
  \bibinfo{pages}{4310} (\bibinfo{year}{1993}), \eprint{hep-ph/9305276}.

\bibitem[{\citenamefont{Blasone and Vitiello}(1995)}]{BlaVit95}
\bibinfo{author}{\bibfnamefont{M.}~\bibnamefont{Blasone}} \bibnamefont{and}
  \bibinfo{author}{\bibfnamefont{G.}~\bibnamefont{Vitiello}},
  \bibinfo{journal}{Ann. Phys.} \textbf{\bibinfo{volume}{244}},
  \bibinfo{pages}{283} (\bibinfo{year}{1995}), \bibinfo{note}{erratum-ibid. 249
  (1996) 363-364}, \eprint{hep-ph/9501263}.

\bibitem[{\citenamefont{Grimus and Stockinger}(1996)}]{GriSto96}
\bibinfo{author}{\bibfnamefont{W.}~\bibnamefont{Grimus}} \bibnamefont{and}
  \bibinfo{author}{\bibfnamefont{P.}~\bibnamefont{Stockinger}},
  \bibinfo{journal}{Phys. Rev. D} \textbf{\bibinfo{volume}{54}},
  \bibinfo{pages}{3414} (\bibinfo{year}{1996}), \eprint{hep-ph/9603430}.

\bibitem[{\citenamefont{Beuthe}(2002)}]{Beu02}
\bibinfo{author}{\bibfnamefont{M.}~\bibnamefont{Beuthe}},
  \bibinfo{journal}{Phys. Rev. D} \textbf{\bibinfo{volume}{66}},
  \bibinfo{pages}{013003} (\bibinfo{year}{2002}), \eprint{hep-ph/0202068}.

\bibitem[{\citenamefont{Dvornikov}(2005)}]{Dvo05}
\bibinfo{author}{\bibfnamefont{M.}~\bibnamefont{Dvornikov}},
  \bibinfo{journal}{Phys. Lett. B} \textbf{\bibinfo{volume}{610}},
  \bibinfo{pages}{262} (\bibinfo{year}{2005}), \eprint{hep-ph/0411101}.

\bibitem[{\citenamefont{Wolfenstein}(1978)}]{Wol78}
\bibinfo{author}{\bibfnamefont{L.}~\bibnamefont{Wolfenstein}},
  \bibinfo{journal}{Phys.\ Rev.\ D} \textbf{\bibinfo{volume}{17}},
  \bibinfo{pages}{2369} (\bibinfo{year}{1978}).

\bibitem[{\citenamefont{Mikheev and Smirnov}(1985)}]{MikSmi85eng}
\bibinfo{author}{\bibfnamefont{S.~P.} \bibnamefont{Mikheev}} \bibnamefont{and}
  \bibinfo{author}{\bibfnamefont{A.~{\relax Yu}.} \bibnamefont{Smirnov}},
  \bibinfo{journal}{Sov. J. Nucl. Phys.} \textbf{\bibinfo{volume}{42}},
  \bibinfo{pages}{913} (\bibinfo{year}{1985}).

\bibitem[{\citenamefont{Mannheim}(1988)}]{Man88}
\bibinfo{author}{\bibfnamefont{P.~D.} \bibnamefont{Mannheim}},
  \bibinfo{journal}{Phys. Rev. D} \textbf{\bibinfo{volume}{37}},
  \bibinfo{pages}{1935} (\bibinfo{year}{1988}).

\bibitem[{\citenamefont{Pantaleone}(1992)}]{Pan92}
\bibinfo{author}{\bibfnamefont{J.}~\bibnamefont{Pantaleone}},
  \bibinfo{journal}{Phys. Rev. D} \textbf{\bibinfo{volume}{46}},
  \bibinfo{pages}{510} (\bibinfo{year}{1992}).

\bibitem[{\citenamefont{Cardall and Chung}(1999)}]{CarChu99}
\bibinfo{author}{\bibfnamefont{C.~Y.} \bibnamefont{Cardall}} \bibnamefont{and}
  \bibinfo{author}{\bibfnamefont{D.~J.~H.} \bibnamefont{Chung}},
  \bibinfo{journal}{Phys. Rev. D} \textbf{\bibinfo{volume}{60}},
  \bibinfo{pages}{073012} (\bibinfo{year}{1999}).

\bibitem[{\citenamefont{Nunokava et~al.}(1997)\citenamefont{Nunokava, Semikoz,
  Smirnov, and Valle}}]{NunSemSmiVal97}
\bibinfo{author}{\bibfnamefont{H.}~\bibnamefont{Nunokava}},
  \bibinfo{author}{\bibfnamefont{V.~B.} \bibnamefont{Semikoz}},
  \bibinfo{author}{\bibfnamefont{A.~{\relax Yu}.} \bibnamefont{Smirnov}},
  \bibnamefont{and} \bibinfo{author}{\bibfnamefont{J.~W.~F.}
  \bibnamefont{Valle}}, \bibinfo{journal}{Nucl.\ Phys.\ B}
  \textbf{\bibinfo{volume}{501}}, \bibinfo{pages}{17} (\bibinfo{year}{1997}),
  \eprint{hep-ph/9701420}.

\bibitem[{\citenamefont{Grigoriev et~al.}(2002)\citenamefont{Grigoriev,
  Lobanov, and Studenikin}}]{GriLobStu02}
\bibinfo{author}{\bibfnamefont{A.}~\bibnamefont{Grigoriev}},
  \bibinfo{author}{\bibfnamefont{A.}~\bibnamefont{Lobanov}}, \bibnamefont{and}
  \bibinfo{author}{\bibfnamefont{A.}~\bibnamefont{Studenikin}},
  \bibinfo{journal}{Phys.\ Lett.\ B} \textbf{\bibinfo{volume}{535}},
  \bibinfo{pages}{187} (\bibinfo{year}{2002}), \eprint{hep-ph/0202276}.

\bibitem[{\citenamefont{Beuthe}(2003)}]{Beu03}
\bibinfo{author}{\bibfnamefont{M.}~\bibnamefont{Beuthe}},
  \bibinfo{journal}{Phys. Rept.} \textbf{\bibinfo{volume}{375}},
  \bibinfo{pages}{105} (\bibinfo{year}{2003}), \eprint{hep-ph/0109119}.

\bibitem[{\citenamefont{Nishi}(2006)}]{Nis05}
\bibinfo{author}{\bibfnamefont{C.~C.} \bibnamefont{Nishi}},
  \bibinfo{journal}{Phys. Rev. D} \textbf{\bibinfo{volume}{73}},
  \bibinfo{pages}{053013} (\bibinfo{year}{2006}), \eprint{hep-ph/0506109}.

\bibitem[{\citenamefont{Lobanov and Studenikin}(2001)}]{LobStu01}
\bibinfo{author}{\bibfnamefont{A.~E.} \bibnamefont{Lobanov}} \bibnamefont{and}
  \bibinfo{author}{\bibfnamefont{A.~I.} \bibnamefont{Studenikin}},
  \bibinfo{journal}{Phys. Lett. B} \textbf{\bibinfo{volume}{515}},
  \bibinfo{pages}{94} (\bibinfo{year}{2001}), \eprint{hep-ph/0106101}.

\bibitem[{\citenamefont{Dvornikov and Studenikin}(2002)}]{DvoStu02JHEP}
\bibinfo{author}{\bibfnamefont{M.}~\bibnamefont{Dvornikov}} \bibnamefont{and}
  \bibinfo{author}{\bibfnamefont{A.}~\bibnamefont{Studenikin}},
  \bibinfo{journal}{J. High Energy Phys.} \textbf{\bibinfo{volume}{09}},
  \bibinfo{pages}{016} (\bibinfo{year}{2002}), \eprint{hep-ph/0202113}.

\bibitem[{\citenamefont{Bogoliubov and Shirkov}(1980)}]{BogShi80p607full}
\bibinfo{author}{\bibfnamefont{N.~N.} \bibnamefont{Bogoliubov}}
  \bibnamefont{and} \bibinfo{author}{\bibfnamefont{D.~D.}
  \bibnamefont{Shirkov}}, \emph{\bibinfo{title}{Introduction to the theory of
  quantized fields}} (\bibinfo{publisher}{Wiley}, \bibinfo{address}{New York},
  \bibinfo{year}{1980}), p. \bibinfo{pages}{607}, \bibinfo{edition}{3rd} ed.

\end{thebibliography}

\end{document}